**Title: Analysis of the spontaneous emission limited linewidth of an integrated III-V/SiN laser**

*Weng W. Chow, Yating Wan*, John E. Bowers and Frédéric Grillot*

Weng W. Chow

Sandia National Laboratories, Albuquerque, NM 87185-1086, U.S.A.

Yating Wan, John E. Bowers

Department of Electronic and Computer Engineering, University of California – Santa Barbara, Santa Barbara, California 93106, U.S.A.

Frédéric Grillot

LTCI, Télécom Paris, Institut Polytechnique de Paris, 91120 Palaiseau, France

Email: yatingwan@ucsb.edu

**Abstract:** This paper describes a calculation of the spontaneous emission limited linewidth of a semiconductor laser consisting of hybrid or heterogeneously integrated, silicon and III-V intracavity components. Central to the approach are a) description of the multi-element laser cavity in terms of composite laser/free-space eigenmodes, b) use of multimode laser theory to treat mode competition and multiwave mixing, and c) incorporation of quantum-optical contributions to account for spontaneous emission effects. Application of the model is illustrated for the case of linewidth narrowing in an InAs quantum-dot laser coupled to a high-$Q$ SiN cavity.

## 1. Introduction

For coherent communication and sensing, there is much effort towards developing narrow linewidth semiconductor lasers, beyond what is achievable with single Fabry-Perot or distributed feedback (DFB) resonators. [1] To reduce spectral linewidth, many approaches are being considered, including external cavity, phase-shifted and chirped grating, discrete mode DFB lasers, and fiber lasers. [2] [3] [4] [5] [6] From the active region aspect, InAs/InP quantum dot (QD) DFB lasers, operating with very low population inversion, have achieved spectral linewidth down to 30 kHz. [7] [8] Recently, much progress is reported on modal engineering of a DFB laser, in which light is generated in the III-V material and stored into the low-loss silicon material. This class of hybrid or heterogeneous integrated lasers indicates a promising path forward, where wavelengths down to 1 kHz level have already been demonstrated. [9] The laser structure relies on using a harmonic potential cavity to produce a large quality factor. A similar design has been proposed and a long photon lifetime of $\mathbf{103\ ps}$ was obtained [10]. Further improvement was achieved with a hybrid integrated laser that included a commercial DFB laser butt-coupled to the bus waveguide of the $Si_3N_4$ resonator chip [11]. With a high cavity quality factor $\boldsymbol{Q > 2.5 \times 10^8}$, an electrically pumped integrated laser with a linewidth of $\mathbf{1.2\ Hz}$ was demonstrated.

This paper describes a calculation of the spontaneous emission limited spectrum of a semiconductor laser consisting of III-V and SiN sections. A composite-cavity mode description is used to treat the effects of optical coupling among different intracavity components in an extended cavity. [12] The composite-cavity modes also provide a treatment of outcoupling that is more consistent with mode projections used in laser theory. [13] [14] Active medium contributions are described within the context of multimode semiclassical laser theory, where electron-hole polarization dynamics accounts for both linear gain and carrier-induced refractive index change, as well as nonlinearities giving rise to saturation, mode competition and multiwave mixing. [14] Strictly speaking, the intrinsic linewidth determination requires radiation field quantization. We have taken such an approach in the past and found the numerical evaluation of the two-time field correlations to be very demanding, especially for parametric studies involving complex resonators and accounting for multimode effects. [15] Instead, we extracted from the single-mode quantum-optical derivation, terms arising from spontaneous emission that can be incorporated into the intensity and frequency determining equations in laser theory. Our approach resembles a Langevin description [16] with the added complication of a complex resonator geometry and a more consistent treatment of outcoupling.

Section 2 describes the formulation of the approach. The concept of composite-cavity modes is discussed, along with their advantages, in terms of validity for arbitrary coupling between III-V and SiN sections, and a consistent treatment of laser outcoupling. Also, the section describes the use of composite-cavity modes in a multimode semiconductor-laser theory to account for active region contributions. The section ends with a discussion on the incorporation of spontaneous emission contributions into the laser equations. Section 3 demonstrates the application of the approach to identify and understand the physical mechanisms leading to spectral narrowing in an extended cavity with III-V active and SiN passive sections. Results are discussed, that are from a parametric study involving the lasing linewidth dependences on the SiN cavity $Q$ and the coupling between III-V and SiN sections.

## 2. Theory

### *2.1 Composite-cavity modes*

A difference in our approach is the treatment of the optically coupled III-V and SiN sections as a combined system. This composite-cavity treatment provides a description that is valid for arbitrary coupling (i.e., from completely isolated to totally coupled). Variations of this approach have been used to investigate cleaved-coupled-cavity lasers, photonic-crystal lasers and isolator-free, injection-locked lasers. [17] [18] [12] [19] [20]

Assuming that transverse and longitudinal spatial dependences of the intracavity field may be decoupled, we use the effective index method to reduce to a 1-dimensional geometry. Then, Maxwell's equations become

$$\left(\frac{\partial^2}{\partial z^2}+\frac{n^2(z)}{c^2}\frac{\partial^2}{\partial t^2}-\frac{\alpha(z)}{c}\frac{\partial}{\partial t}\right)E(z,t)=-\mu_0\frac{\partial^2}{\partial t^2}P(z,t)\,, \qquad (1)$$

where $E(z,t)$ is the radiation field, $P(z,t)$ is the polarization representing the active medium, $\alpha(z)$ describes the losses (such as intracavity absorption) at various locations, $\mu_0$ and $c$ are the permeability and speed of light in vacuum. In Equation (1), $t$ is time and $z$ is the position along the III-V and SiN cavity axis. The arrangement of intracavity optical components is described by the z-dependence of the effective refractive index $n(z)$. Setting ð$^2P/\partial t^2=\alpha=0$ and writing:

$$E(z,t)=E_m(t)\cos(\nu_m t)u_m(z)\;, \qquad (2)$$

we obtain the equation for a composite passive cavity eigenmode:

$$\frac{d^2}{dz^2}u_m(z)=\frac{n^2(z)}{c^2}\Omega_m^2u_m(z)\;, \qquad (3)$$

where $m$ is the mode index and $\Omega_m$ is the eigenfrequency.

We also use the composite-cavity modes to circumvent an inconsistency in laser physics involving the treatment of the active medium and optical resonator contributions. [13] [14] In laser theory, passive-cavity normal modes are necessary for performing the projections leading to the laser field amplitude- and frequency-determining equations. However, the rigorous definition of normal modes for a laser cavity is not possible because of mirror outcoupling. One customarily uses a set of quasi modes obtained from, e.g. a Fox-Li calculation, and accounts for cavity losses by adding a decay term to the amplitude-determining equation. Such a phenomenological approach is inadequate for describing the finer details of laser linewidth, such as the evolution of the emission spectrum during the transition from below to above lasing threshold. Instead, we follow an earlier paper treating outcoupling in a Fabry-Perot laser [21] by extending the effective refractive index $n(z)$ in Equation (3) to include a very long resonator approximating free-space (see **Figure**. 1).

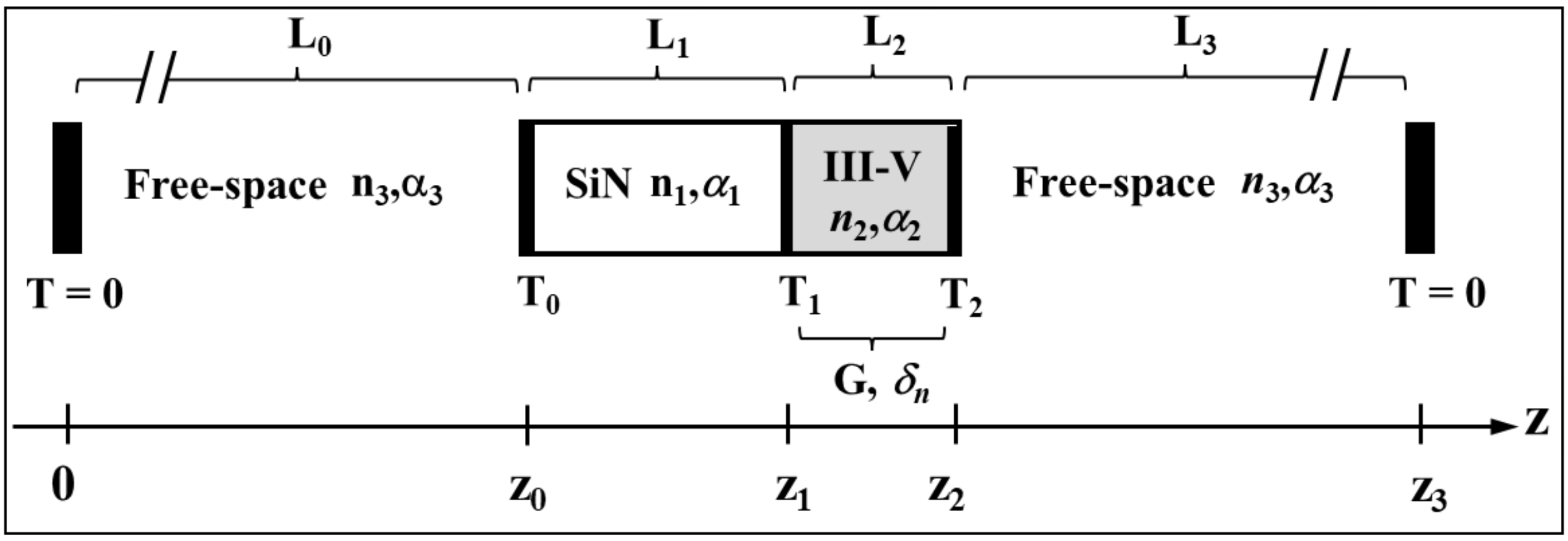

**Figure 1**. Basic III-V/SiN coupled-cavity configuration used in the calculations. Input to the laser theory are the passive cavity resonances. We specify them using the cavity lengths $L_1$ and $L_2$, and transmissions $T_0$, $T_1$ and $T_2$. The transmission $T_1$ also approximates the effective coupling between SiN and III-V cavities. The gain and carrier-induced refractive index ($G$ and $\delta n$, respectively) are calculated from the laser theory. $L_3$ is made sufficiently large to provide sufficient composite-cavity modes to accurately reproduce the finite-width (Fox-Li) resonances caused by outcoupling.

Strictly speaking, Equation (3) should be solved using the precise $n(z)$ describing the spatial variations in effective refractive index from the different material layers making up the distributed Bragg reflectors, spacer layers and waveguides. Such a detailed description is unnecessary as well as cumbersome for our present goal, which is a general understanding of the linewidth results when a laser is coupled to a very high-$Q$ passive resonator. Hence, we will start with the basic 1-d geometry depicted in Figure 1, where the interfaces between optical sections are treated by 'bumps' in $n(z)$. [22] In this case, the composite-cavity modes have the boundary conditions:

$$u_m(0) = u_m(z_3) = 0, \tag{4}$$

$$u_m(z_i^+) = u_m(z_i^-) = 0, \tag{5}$$

$$\frac{d}{dz}u_m(z_i^+) - \frac{d}{dz}u_m(z_i^-) = -\eta_i k u_m(z_i) \ , \tag{6}$$

where

$$\eta_i = 2\sqrt{\frac{(1-T_i)}{T_i}} \ , \tag{7}$$

$z_i^-$ and $z_i^+$ are located immediately prior and after $z_i$ , $T_i$ is the effective transmission representing the coupling between sections and $k$ is the average magnitude of the wave vector. The orthogonality relation comes from integrating by parts Equation (3):

$$\int_0^{z_3} dz\, n^2(z)\, u_n(z) u_m(z) = \delta_{n,m} N_c \ , \tag{8}$$

where $N_c = \sum_{j=0}^{3} n_j L_j$ is the normalization.

To illustrate the basic properties of composite-cavity modes, we consider the example of two cavities of lengths and refractive indices $L_1 = 4\ mm$, $n_1 = 2.1$ and $L_2 = 600\ \mu m$, $n_2 = 3.6$. The cavities are coupled via $T_1 = 0.01$, and the outcoupling is $T_0 = 0$ and $T_2 = 0.02$. **Figure** 2 (a) are plots of the resonances of the uncoupled III-V and SiN cavities, obtained by solving Equation (3). When coupled, the solution to Equation (3) gives the blue and grey curves in Figure 2 (b). They show the resonances in the SiN and III-V cavities, specifically, the overlap of each composite-cavity mode within the two sections of the laser cavity: $\Gamma_{1,m}^{(1)} = N_c^{-1} \int_{z_0}^{z_1} dz\ n_1^2\ u_m^2(z)$ versus $\Omega_m$ and $\Gamma_{2,m}^{(1)} = N_c^{-1} \int_{z_1}^{z_2} dz\ n_2^2\ u_m^2(z)$ versus $\Omega_m$. The magnitude of an integral gives of measure of the faction of the mode residing in a particular section compared to in free space. The fraction, $\Gamma_{2,m}^{(1)}/\left(\Gamma_{1,m}^{(1)} + \Gamma_{2,m}^{(1)}\right)$ is the mode confinement factor typically present a laser gain formula derived with quasi-(Fox-Li) modes.

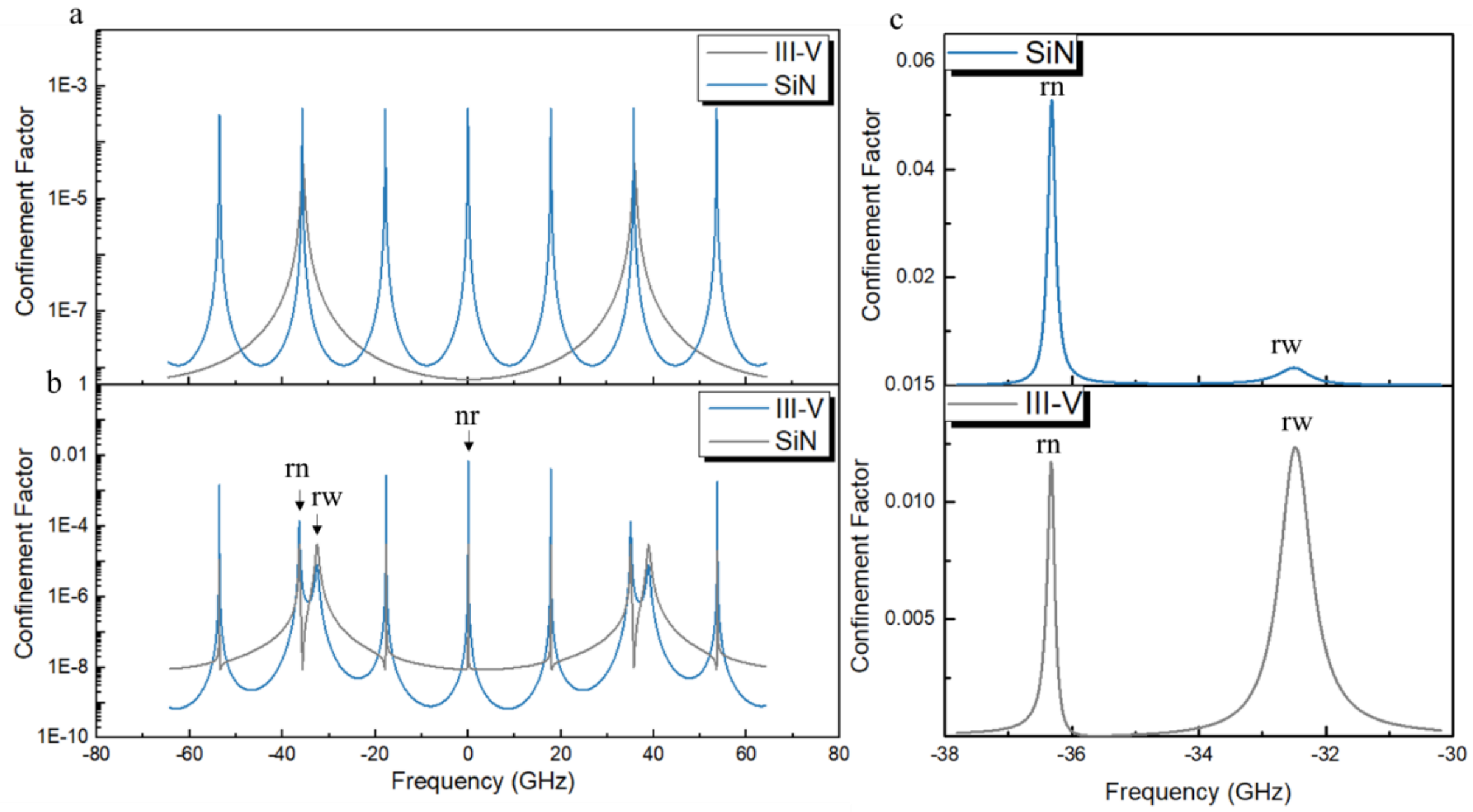

**Figure 2**. (a) Resonances of individual III-V cavity (grey) and SiN cavities (blue). (b) Resonances of the coupled system, in the III-V (grey) and SiN cavities (blue). The arrows label the resonances studied in the next section: resonant narrow (rn), resonant wide (rw) and non-resonant (nr). (c) Higher resolution of double peak resonance within SiN cavity (blue) and III-V cavity (grey). Each composition mode contributes to a point in the resonances.

There are basically 3 kinds of resonances. Where the uncoupled III-V and SiN resonances coincide, Figure 2 (b) shows two closely spaced composite-cavity resonances (rn and rw), with splitting determined by the coupling $T_1$. The other resonances (e.g. nr) belong to the eigenmodes of the longer SiN cavity. They also appear in the III-V cavity (grey curve) because of cavity coupling.

Without coupling to free-space, the resonances in Figure 2 would be delta functions. Outcoupling leads to finite-linewidth resonances (Fox-Li quasi modes) as depicted in Figure 2c for the resonant case. Each resonance is composed of multiple composite-cavity modes as indicated by the dots. Based on linewidths, one may associate the broad resonance (rw) with the III-V cavity, where cavity length $L_2 = 600\ \mu m$ and facet transmissions $T_1 = 0.01, T_2 = 0.02$ result in 550 MHz full-width at half-maximum (FWHM) and $Q = 3.9 \times 10^5$. The narrow resonance (rn) may be assigned to the SiN cavity, with length $L_1 = 4\, mm$ and facet transmissions $T_0 = 0$, $T_1 = 0.01$ giving $FWHM = 60\, MHz$ and $Q = 3.6 \times 10^6$. The blue curve in Figure 2(c) for the SiN cavity also shows a weak, broad resonance injected from the III-V cavity. Similarly, the grey curve in Figure 2 (c) for the III-V cavity shows a narrow resonance injected from the SiN cavity. The difference in injection levels between the two cavities is because the SiN cavity high $Q$ inhibits external influences, whereas the III-V cavity lower $Q$ is less discriminating. Hence the non-reciprocity in mutual injection or self-feedback.

### 2.2 Laser theory

To include the InAs quantum-dot active medium, we follow semiclassical laser theory [14] and write

$$P(z,t) = \frac{1}{2} e^{-i\Omega_0 t} \sum_n P_n(t) e^{-i\phi_n(t)} u_n(z) + c.c.\ , \tag{9}$$

where $P_n(t)$ is the complex polarization amplitude that connects Equation (1) to a quantum mechanical electron-hole polarization $p_q(t) = \langle b_q c_q \rangle \exp[(i\omega_q + \gamma)t]$ via

$$P_n(t) = \frac{2\Gamma_{nn}^{(1)} \wp}{V} e^{i\nu_n t + i\phi_n(t)} \frac{1}{N_c} \int_{z_1}^{z_2} dz\, n^2(z)\, u_n(z) \sum_q p_q(t)\ , \tag{10}$$

where we assumed that the gain region extends over the entire III-V cavity length. In the above equation, $\Gamma_{nn}^{(1)}$ is the mode confinement factor for the gain region, $\gamma$ is the dephasing rate, $\wp$ is the dipole matrix element for the interaction between an electron-hole pair and the laser field, and $c_q$ and $b_q$ are electron and hole annihilation operators, respectively. For narrow linewidth lasers, where lasing should involve only the ground-state transition, we use $q$ to label a group of QDs with the same ground state transition frequency $\omega_q$. The $q$ summation is over the inhomogeneous QD distribution.

The laser derivation gives the intensity- and frequency-determining equations for each composite laser/free-space mode,

$$\frac{dI_n}{dt} = \left[g_n^{sat}\left(N^{(2d)}\right) - \gamma_n^{cav}\right]I_n + S_n\left(N^{(2d)}\right)$$

$$+ \sum_{m\neq n} 2\sqrt{I_n I_m}\mathrm{Re}\left[B_{nm}\left(N^{(2d)}\right)e^{-i\psi_{nm}}\right], \qquad (11)$$

$$\frac{d\psi_n}{dt} = \Omega_n + \left[\sigma_n\left(N^{(2d)}\right) - \sum_m \tau_{nm}\left(N^{(2d)}\right)I_m\right] + S_n^{\phi}\left(N^{(2d)}\right)$$

$$+ \sum_{m\neq n} \sqrt{\frac{I_m}{I_n}}\mathrm{Im}\left[B_{nm}\left(N^{(2d)}\right)e^{-i\psi_{nm}}\right], \qquad (12)$$

where $d\psi_n/dt = \nu_n + d\phi_n/dt$ is the n$^{th}$ composite-cavity mode lasing frequency, and we define $\psi_{nm} = \psi_n - \psi_m$ and $I_n = (\wp E_n/(2\hbar\gamma))^2$. In the right-hand side of Equation (11), the first two terms are the modal saturated gain and cavity loss, the third term accounts for the spontaneous emission and the last term is from the first order polarization, arising because the composite-cavity modes are not orthogonal when integrated over only the gain region. The square bracket in Eqs. (12) contains the modification to the passive composite-cavity mode frequency $\Omega_n$ by the active medium. In laser theory, they are referred to as frequency pulling and pushing $\sigma_n$ and $\tau_{nm}$, respectively. In semiconductor laser models, they describe the carrier-induced refractive index change and is typically taken into account via the linewidth enhancement factor. [23]

In our treatment, all the coefficients associated in Eqs. (11) and (12) are derived from the electron-hole polarization equation of motion, and are therefore, calculated instead of fitting parameters (see equations in Table 1). More details on the active medium contributions are discussed later, when we explain the line narrowing mechanisms within the context of a laser/free-space composite-cavity mode picture. A difference between a semiconductor laser and an atomic or molecular one is that it cannot be categorized as either homogeneously or inhomogeneously broadened. Owing to rapid carrier-carrier scattering, a semiconductor laser tunes inhomogeneously and saturates homogeneoulsy. [24] A consequence is that the intensity- and frequency-determining equations alone do not determine laser behavior. An expedient approach is to evaluate all active medium coefficients at the saturated carrier density $N_{2d}$. To do this, Eqs. (11) and (12) are solved simultaneously with the total carrier density equation of motion:

$$\frac{dN_{2d}}{dt} = \frac{\epsilon_B h_{qw}}{8\hbar\nu_0}\left(\frac{\wp}{2\hbar\gamma}\right)^2 \frac{1}{\Gamma_{xy}}\sum_n g_n^{sat} I_n$$

$$+\frac{\eta_p J}{eN_{qw}} - \gamma_{nr}N_{2d} - B_{spont}^{(2d)}N_{2d}^2 , \qquad (13)$$

where $\epsilon_B$ and $\nu_0$ are the averaged permittivity and frequency, $h_{qw}$ is the thicknesss of one of the quantum wells and $\Gamma_{xy}$ is the transverse confinment factor. For the injection current contributions, $\eta_p$ is the pump efficiency, $J$ is the injection current density, $N_{qw}$ is the number of quantum wells in the active medium. For carrier losses, we use the effective rates, $\gamma_{nr}$ for the nonradiative (Schottky-Read-Hall) carrier loss and $B_{spont}^{(2d)}$ for the bimolecular carrier recombination due to spontaneous emission. While there are more rigorous treatments, our approach has the advantage of reducing numerical demands and yet remaining connected to microscopic details such as bandstructure. Its accuracy depends on laser operation close to quasi-equilibrium condition, which is likely satisfied during narrow linewidth operation.

**Table 1**. Active medium coefficients, where $L_\gamma(x) = 1/[1 + (x/\gamma)^2]$, $F_1 = \wp^2\nu_0 N_{QD}^{(2d)}/(2\hbar\gamma\epsilon_B h_{qw})$ and $D_y(x) = 1/(1 + i\,x/y)$

| Parameter | Equation |
|---|---|
| Saturated gain | $g_n^{sat} = g_n/\left(1 + \sum_m \kappa_{nm} I_m\right)$ |
| Linear gain and frequency pulling | $g_n/2 + i\sigma_n = F_1\Gamma_{xy}\Gamma_{nn}^{(1)}\Lambda_n^{(1)}N_{inv}$ |
| Inversion | $N_{inv} = f(\varepsilon_q^e, \mu_e, T) + f(\varepsilon_q^h, \mu_h, T) - 1$ |
| Frequency locking | $B_{nm} = F_1\Gamma_{xy}\Gamma_{nm}^{(1)}\Lambda_m^{(1)}N_{inv}$ |
| Gain compression | $\kappa_{nm} = 2\Gamma_{nm}^{(3)}\gamma\mathrm{Re}(\Lambda_{nm}^{(3)})/[\Gamma_{nn}^{(1)}\gamma_{ab}\mathrm{Re}(\Lambda_{nn}^{(1)})]$ |
| Frequency pushing | $\tau_{nm} = 2F_1\Gamma_{nm}^{(3)}\mathrm{Im}(\Lambda_{nm}^{(3)})\gamma/\gamma_{ab}$ |
| Linear mode confinement | $\Gamma_{nn}^{(1)} = N_c^{-1}\int_{L_1}^{L_1+L_2} dz\, n^2(z)u_n(z)u_m(z)$ |
| Nonlinear mode confinement | $\Gamma_{nm}^{(3)} = N_c^{-1}\int_{L_1}^{L_1+L_2} dz\, n^2(z)u_n^2(z)u_m^2(z)$ |
| Linear susceptibility spectral contribution | $\Lambda_n^{(1)} = \sum_q D_\gamma(\Delta_{nq})$, where $\Delta_{nq} = \nu_n - \omega_q$ |
| Nonlinear suceptibility spectral contribution | $\Lambda_{nn}^{(3)} = \sum_q D_\gamma(\Delta_{nq})L_\gamma(\Delta_{nq})$<br>$\Lambda_{nm}^{(3)} = \gamma/\gamma_{ab}\sum_q D_\gamma(\Delta_{nq})\{2L_\gamma(\Delta_{mq})$<br>$+D_{\gamma_{ab}}(\nu_n - \nu_m)[D_\gamma(\Delta_{nq}) + D_\gamma(\Delta_{qm})]\}$ |

## 2.3 Connection to quantum optics

We close this section with an explanation of how we incorporate the effects of spontaneous emission into our basically semiclassical laser model. From a cavity-QED derivation, we obtain the equation of motion for the single-mode intracavity photon number. Assuming that the electron-hole polarization changes sufficiently fast to follow any time variation in the photon and carrier populations (rate equation approximation and consistent with quasi-equilibrium condition), the polarization may be adiabatically eliminated. In the resulting photon number equation of motion, the spontaneous emission contribution appears explicitely as a bimolecular carrier recombination term. [25] Using the conversion,

$$I_n = \left(\frac{\wp}{2\hbar\gamma}\right)^2 \frac{\hbar\nu_n}{\epsilon_B V_{mode}} n_n \tag{14}$$

we obtain

$$S_n = \frac{\varepsilon_{g0} N_{qw} w L_g}{\epsilon_B V_{mode}} \left(\frac{\wp}{2\hbar\gamma}\right)^2 \Gamma_{nn}^{(1)} \beta_{spont} B_{spont}^{(2d)} N_{2d}^2 f(\varepsilon_n^e, \mu_e, T) f(\varepsilon_n^h, \mu_h, T)\,, \tag{15}$$

where $n_n$ is the photon number in the n^th^ composite-cavity mode, $V_{mode}$ is the composite-mode volume, $w$ and $L_g$ are the stripe width and length of the active region, $\beta_{spont}$ is the spontaneous emission factor, $f(\varepsilon_n^e, \mu_e, T)$ and $f(\varepsilon_n^h, \mu_h, T)$ are the electron and hole populations (assuming Fermi functions) contributing to the spontaneous emission.

To obtain the spontaneous emission contribution to the frequency determining equation Equation (12), we derive the equaton of motion for the photon annihilation operator in a composite-cavity mode. Working in the Interaction Picture, the equation of motion for the photon annihilation operator in a composite-cavity mode, to 3$^{rd}$ order in light-matter interaction, is [14]

$$\frac{dA}{dt} = \left(g - \frac{\gamma_{cav}}{2}\right) A - \beta A A^{\dagger} A + G\,, \tag{16}$$

where $g$ and $\beta$ are the linear and nonlinear amplitude gain coefficients, $G$ is the Langevin force operator from spontaneous emission and for brevity, we drop the mode index. Assuming that photon number fluctuation is negilible and $\langle A(0)\rangle = \sqrt{n_p}$, we write for the slowly varying photon annihilation operator,

$$\begin{aligned} \langle A(t)\rangle &= \sqrt{n_p}\langle \exp[-i\phi(t)]\rangle\,, \\ &= \sqrt{n_p}\,\langle 1 + i\phi(t) - \frac{1}{2}\phi(t)\phi(t) + \cdots \rangle\,, \\ &= \sqrt{n_p}\exp[-\langle \phi(t)\phi(t)\rangle/2] \end{aligned} \tag{17}$$

Recalling that we neglect photon number fluctuation, we obtain from Equation (16) the quantum optical contribution to the frequency-determining equation:

$$\frac{d\phi(t)}{dt}=\frac{i}{2\sqrt{n_p}}\left[G(t)e^{i\phi(t)}-G^{\dagger}(t)e^{-i\phi(t)}\right] \quad , \tag{18}$$

which we formally integrate and perform more algebra to get

$$\langle\phi(t)\phi(t)\rangle=\frac{1}{4n_p}\int_0^t dt_1\int_0^t dt_2\,\Big\langle G^{\dagger}(t_1)e^{-i\phi(t_1)}G(t_2)e^{i\phi(t_2)} + G(t_1)e^{i\phi(t_1)}G^{\dagger}(t_2)e^{-i\phi(t_2)}\Big\rangle \tag{19}$$

For the correlations involving the Langevin force operators, we perform a separate lengthy calculation, where we make use of Einstein relation [16] [26], to obtain:

$$\langle G^{\dagger}(t)G(t')\rangle+\langle G(t')G^{\dagger}(t)\rangle=2\gamma_{cav}\delta(t-t') \ . \tag{20}$$

Equations (19) and (20) gives $\langle\phi(t)\phi(t)\rangle=\gamma_{cav}t/\left(2n_p\right)$. Hence, to account for phase diffusion due to spontaneous emission, we add to the semiclassical frequency determining equation an imaginary quantity,

$$S_n^{\phi}=i\gamma_n^{cav}\frac{\epsilon_B V_{mode}}{2\hbar\nu_n}\left(\frac{\wp}{2\hbar\gamma}\right)^2\frac{1}{I_n} \tag{21}$$

## 3. Intrinsic linewidth of integrated III-V/SiN laser

This section describes application of the approach in Sec. 2 to study the physical processes occurring when a laser is coupled to a high-$Q$ passive resonator. We consider laser configurations giving the composite-cavity spectra such as in Figure 2 (b). Furthermore, we assume that an intracavity filter, such as a DBR, is present to enable single quasi (Fox-Li) mode operation. For the active region, we use a design that has performed very well in experiments. [27] It consists of 5 $In_{0.15}Ga_{0.85}As$ QWs, each $7\,nm$ thick and embedding a density of $2\times10^{10}\ cm^{-2}$ InAs QDs. From electronic structure calculations, the dipole matrix element $\wp=e\times0.6nm$ and ground-state transition energy is $\hbar\omega_0=0.943\,eV$. Base on quantum kinetic calculations and single-section laser measurements, [28] we use $\gamma=2\times10^{12}s^{-1}$ for the dephasing rate, $\gamma_{ab}=10^{11}s^{-1}$ and $\gamma_{nr}=10^{9}s^{-1}$ for the inter-QD population relaxation and nonradiative decay rates, respectively, carrier injection efficiency $\eta_p=0.35$ and QD inhomogeneous broadening $\Delta_{inh}=15\ meV$. [29] [30].

To obtain the results in this section, we first solve Equation (3) with boundary conditions Eqs. (4) – (6) for the composite-cavity mode frequencies and eigenfunctions. Using the latter, we compute the mode confinement and overlap factors $\Gamma_{nm}^{(1)}$ and $\Gamma_{nm}^{(3)}$ for evaluating the laser coefficients in Table 1. A typical quasi-mode (Fox-Li) resonance consists of a few hundred composite-cavity modes. The mode intensities and lasing frequencies are determined by intensity- and frequency-determining Eqs. (11) and (12). We numerically solve the coupled Eqs. (11) - (13) until steady state is reached and compute the output power and lasing spectrum for a given current:

$$P = \gamma_{out} \frac{1}{4} \varepsilon_B V_{mode} \left(\frac{\wp}{2\hbar\gamma}\right)^2 \sum_n \Lambda_n^{(1)} I_n \tag{22}$$

$$S(\omega) = \frac{1}{T} \int_0^{t+T} dt_1 \int_0^{\infty} d\tau \, E_{tot}(t_1 + \tau)\, E_{tot}(t_1) e^{i\omega\tau} \tag{23}$$

where

$$E_{tot}(t) = \sum_n \sqrt{I_n(t)}\, exp[i\psi_n(t)] \tag{24}$$

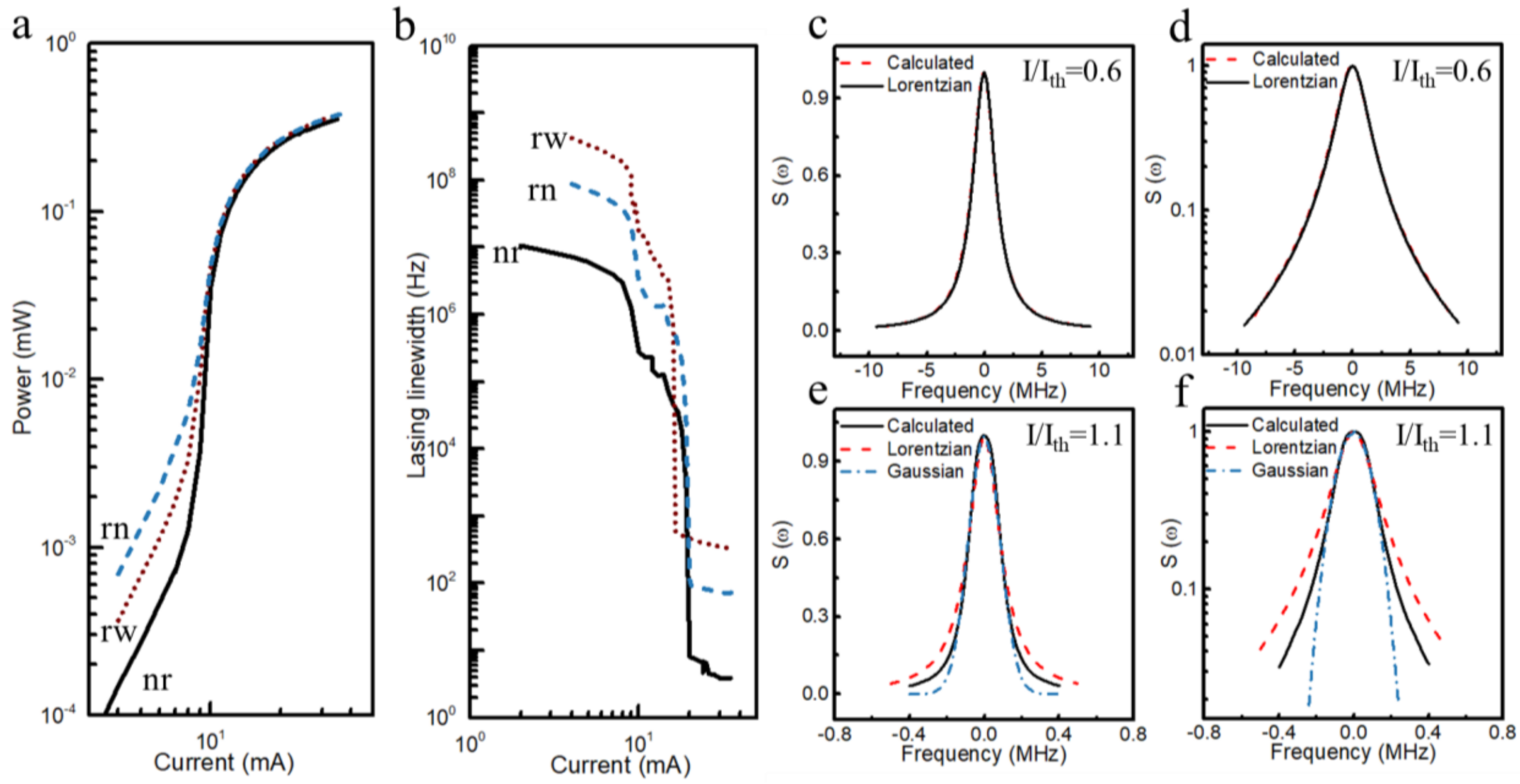

**Figure 3**. Calculated (a) output power and (b) linewidth (full width at half maximum) versus injection current. The effective coupling between III-V and SiN sections is $T_1 = 0.02$ and the curves are for the composite-cavity modes (nr, rn and rw) as indicated in Figure 2 (b). Emission spectra from III-V/SiN laser below ((c) and (d)) and above ((e) and (f)) lasing threshold. The black curves are calculated, the red dashed and blue dash-dotted curves are the fits using Lorentzian and Gaussian functions, respectively. The fits to the main body of the spectrum are performed using (c) and (e). (d) and (f) show the spectral tails.

**Figure 3** shows the computed injection current dependences of output power and lasing linewidth for three composite-cavity modes. The light-current (*L-I*) curves are essentially similar for both resonant and non-resonant configurations. Displacements of the linewidth curves are according to the width of the passive composite-cavity mode resonances. However, their shapes are similar and indicate that spectral narrowing occurs in two stages. From the onset of lasing to twice the threshold, there is already appreciable stimulated emission occurring to produce noticeable narrowing of the spontaneous emission spectrum. The linewidth decreases from the passive cavity value with increasing intracavity intensity from gain clamping or narrowing, as in single-Fabry Perot or distributed feedback (DFB) lasers. At excitations above twice the threshold, a behavior unique to coupled cavities comes into play. There is a drastic further decrease in linewidth caused by frequency locking of composite-cavity modes. The mechanism involves the term containing the relative phase $\psi_{nm}$ in Equation (12). With higher injection current, the linewidth settles to a value solely determined by $S_n^{\phi}$ in Equation (21).

The simulations uncover two interesting features of the emission spectra during the transition from below to above lasing threshold. One is a change in spectral shape. Below threshold, the entire amplified spontaneous emission spectrum is exactly a Lorentzian function (Figure 3 (c) and 5 (d)). Above the lasing threshold, the lineshape deviates from the Lorentzian function (compare black solid and red dashed curves in Figure 3 (e) and (f)). The blue dotted curves suggest a better fit to a Gaussian function down to about 20% of the spectral peak. This lineshape change has been reported earlier. [31] [32] A quantum optical calculation using a quasi-mode description of outcoupling, is unable to describe the new shape. In the present picture, the slightly flatter shape close to the spectral peak and the sharper drop at the spectral tails come from partial locking of the composite-cavity modes.

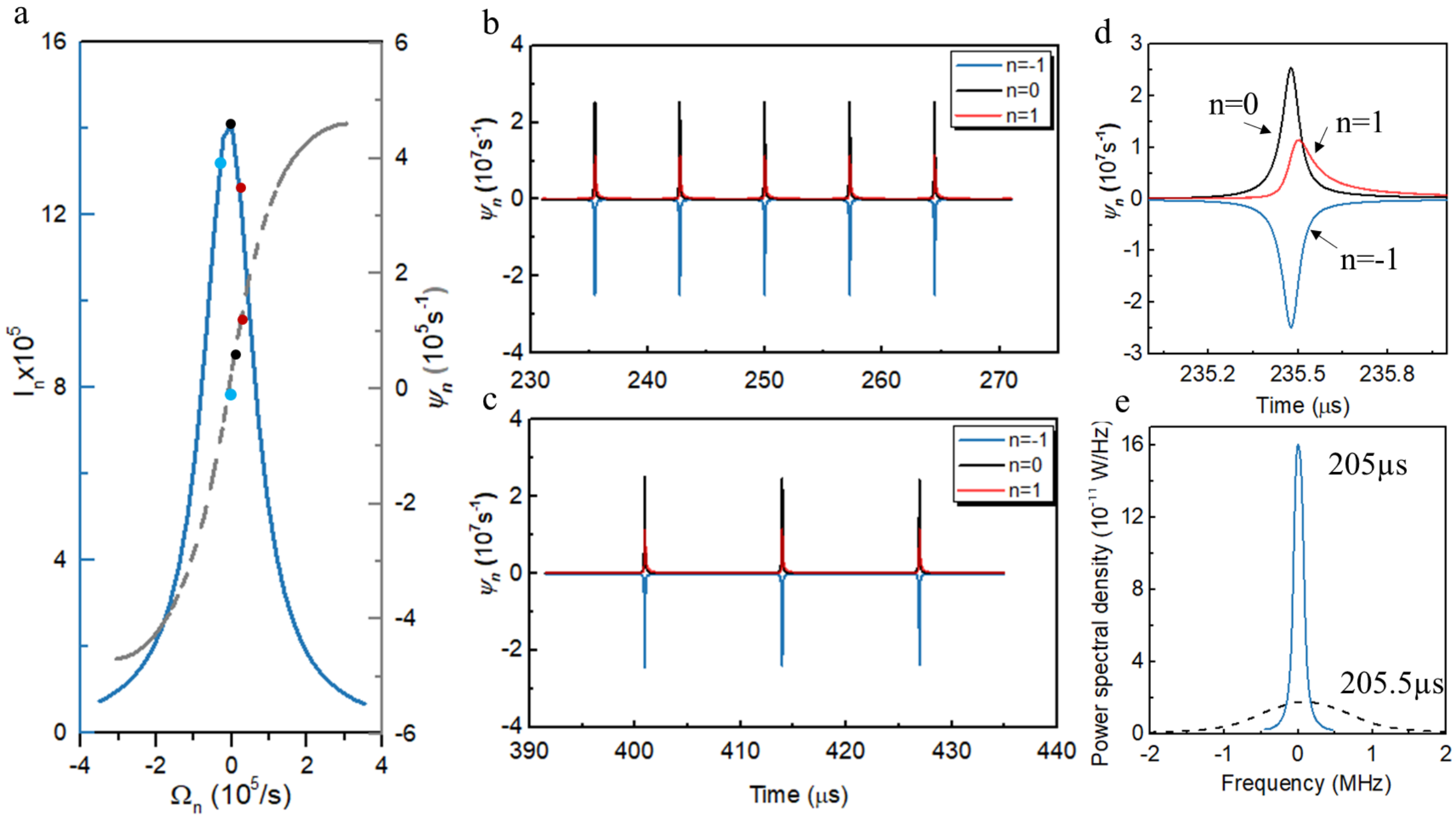

**Figure 4**. (a) Modal intensity and lasing frequencies (solid and dashed curves, respectively) versus passive composite-cavity frequency after steady-state is reached with an injection current of 10 mA. The points indicate

the modal frequencies whose time dependences are tracked. We label those modes (from left to right) $n = -1$, 0 and 1. The time dependences of the lasing frequencies of tracked modes at currents of (b) 10 mA and (c) 15 mA. (d) A higher-time resolution of the modal lasing frequencies in the neighborhood of the spikes for $I/I_{th} = 1.1$ . (e) The lasing spectra for the times as indicated. On the x-axis, frequency is referenced to the spectral peak at 205 μs. The entire spectrum at 202.5 μs is shifted to allow plotting on the same figure.

The partially-locked regime is the second interesting feature. For a closer examination, we monitor time dependences of the lasing frequencies, $d\psi_n/dt$ for the three composite-cavity modes indicated in **Figure 4** (a). In the absence of locking, $d\psi_n/dt$ versus $\Omega_n$ is a straight line with a slope of unity. Deviation from that, as depicted by the dashed curve in Figure 4 (a), indicates the onset of frequency locking. When completely locked, $d\psi_n/dt$ versus $\Omega_n$ is a straight line with zero slope. Figure 4 (b) shows that the lasing frequencies are mostly tightly bunched, except when interrupted by spikes. Earlier reports have associated the spikes with dark solitons. [33] [34] Comparison of Figure. 4 (b) and 4 (c) for $I/I_{th} = 1.1$ and $I/I_{th} = 1.7$, shows an increased period between spikes. This results in narrower average linewidth from the $dt_1$ integration in Equation (23). For $I/I_{th} > 2.2$, stable, complete locking occurs, the spikes vanish and one has the narrowest achievable lasing spectrum that reverts to a Lorentzian lineshape.

In the composite-cavity mode picture, the spikes are instances of coherence collapse when the system breaks lock. Figure 4 (d) shows the coherence collapse region in greater detail. The $n = 0$ trace indicates a blue shift of the entire spectrum by roughly 4 MHz, and the greater separation between traces indicates significant spectral broadening. This is clearly evident in Figure 4 (e) where we plot the spectra between spikes and close to the spike maximum (solid and dashed curves, respectively). The coherence collapse behavior may be described by Alder's equation, $d\psi/dt = a + b\sin(\psi)$, when $b \gtrsim a$, i.e. operation just outside the lockband. As $b$ increases, the duration between spikes increases as shown in Figure. 4 (b) and 4 (c). The duration of each spike also decreases. The average spectral width (averaged over several pike periods) then further decreases, until complete frequency locking is reached and coherence collapse disappears completely.

When formulating our theory, we made sure that it is suitable for parametric studies to produce timely results that are useful for engineering design. We will demonstrate this by presenting results from a parametric study on the effects of optical coupling between III-V and SiN sections. The experimental implementations of coupling include quantum-well intermixing and evanescent field coupling. For the present study, we use the effective transmission $T_1$ as a measure of the net coupling effect. As in an experiment, varying the optical coupling also alters the passive cavity linewidths, the most important being that of the high-Q SiN cavity. In the parametric study, we vary $T_1$ to investigate the competing effects of cavity-$Q$ increase and optical coupling decrease with decreasing optical coupling.

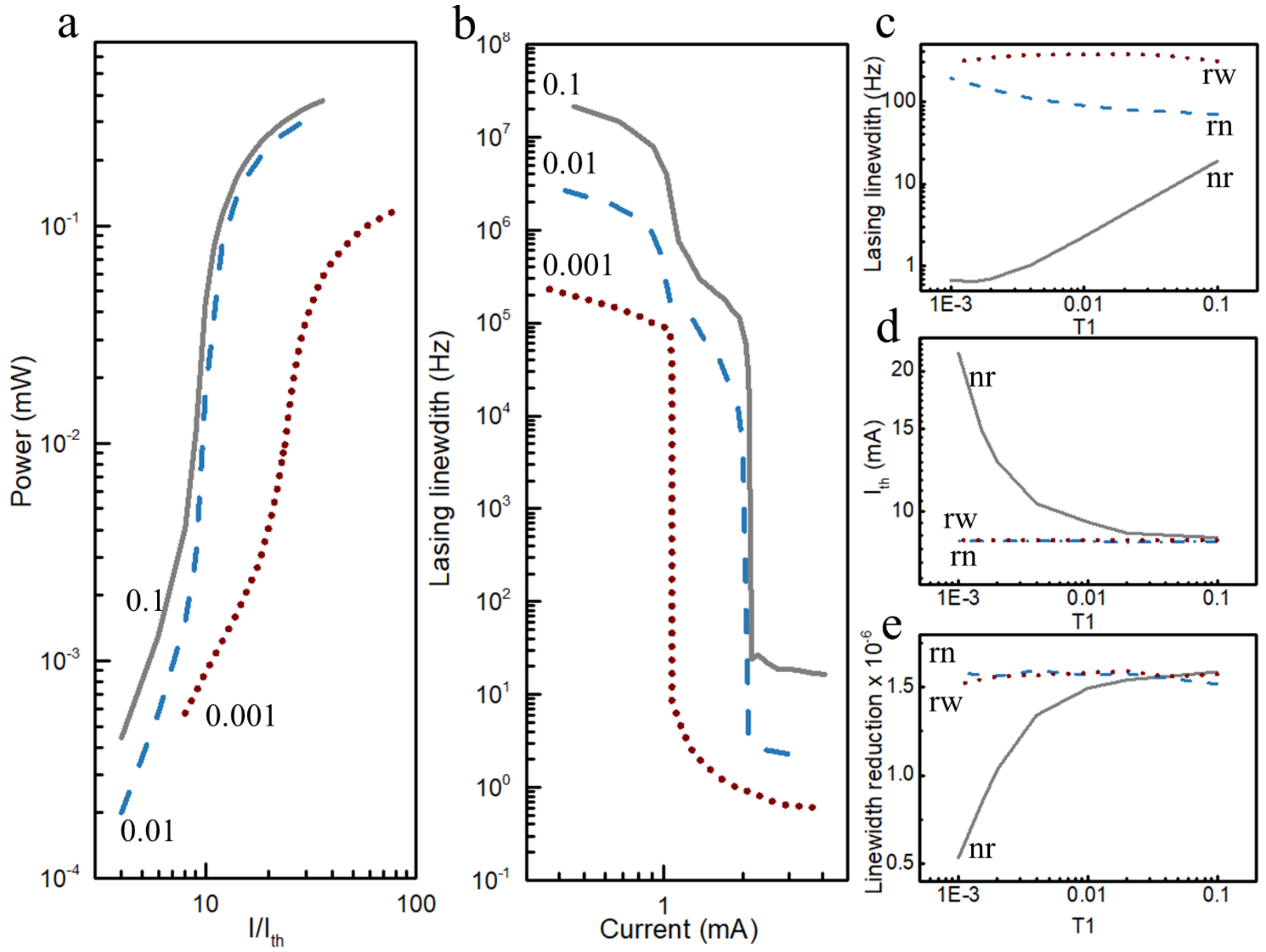

**Figure 5**. (a) *L*-*I* and (b) FWHM versus excitation for different effective coupling between III-V and SiN sections. The curves are for non-resonant (nr) configuration. (c) Lasing spectrum FWHM and (d) threshold current versus coupling between III-V and SiN sections. $T_1$ is an effective transmission representing the interfacial or evanescant coupling between III-V laser and high-$Q$ SiN cavity. The linewidths in Figure 5 (c) are for $I/I_{th} = 3$, where the composite-cavity modes are fully locked. (e) Lasing linewidth reduction versus III-V/SiN optical coupling. The curves are for lasing at the non-resonant (nr) and resonant (rn and rw) configurations.

**Figure 5** (a) and (b) shows the results for operating with different optical coupling for the non-resonant (nr) mode. Simulations indicate little change in *L*-*I* behavior until $T_1$ reduces below $0.01$. After which, mode confinement factor reduction leads to significant degradation in *L*-*I* behavior, as shown in Figure 5 (a) by the red dotted curve for $T_1 = 0.001$. In Figure 5 (b), we plot the FWHM of the lasing spectrum versus injection current relative to the threshold value. The vertical displacement of the curves indicates the linewidth narrowing from decreasing passive SiN cavity $Q$. For $T_1 \geq 0.01$, spectral narrowing takes place in two stages as discussed earlier. With further decrease in coupling, there is a gradual merging of the gain narrowing and frequency locking stages, that eventually results in a single abrupt drop in lasing linewidth as depicted by the $T_1 = 0.001$ curve.

The above calculations are repeated for the resonant modes (rn and rw). We operate the laser at $I/I_{th} = 3$, which is sufficiently high to fully frequency lock the lasing composite-cavity modes, thus giving the minimum possible linewidth. For the non-resonant mode (nr), Figure 5 (c) indicates a decreasing laser linewidth with decreasing $T_1$, because of increasing SiN cavity $Q$, consistent with Figure 5 (b). At $T_1 = 0.002$, the linewidth narrowing saturates, when the increase in SiN cavity $Q$ is balanced by the decrease in III-V/SiN coupling. Figure 5 (c) also indicates that the dependence of linewidth on III-V/SiN optical coupling is vanishing small, in comparison, for the resonant modes (rn and rw). This is because two identical oscillators are always strongly coupled regardless of the coupling. These differences in optical coupling dependences also translate to the threshold current, as shown in Figure 5 (d).

Figure 5e summarizes the combined effects of cavity Q and optical coupling on the minimum achievable laser linewidth. Plotted is the linewidth reduction (SiN cavity linewidth / laser linewidth) versus optical coupling for resonant and non-resonant operation. The calculations are performed for 3 times threshold current to ensure complete locking of the lasing composite-cavity modes, so as to have the maximum linewidth reduction. For the non-resonant (nr) case, the simulations show a relatively constant, large reduction of ~$1.5 \times 10^6$ for $T_1 > 0.01$ (black solid curve). For smaller $T_1$, the curve shows the detrimental effect of too little III-V/SiN coupling, resulting in a sharp drop in linewidth reduction. The curves for the resonant (rn and rw) cases show independence of linewidth reduction on III-V/SiN optical coupling. The reason is that resonant operation is always strongly coupled regardless of optical coupling.

## 1. Conclusion

This paper describes a theoretical approach to investigate the linewidth narrowing when a semiconductor quantum-dot laser is optically coupled to a high-$Q$ silicon nitride resonator. The approach uses composite-cavity eigenmodes to treat the III-V laser, passive SiN resonator and free-space as one combined system. There are two advantages. One is validity for arbitrary optical coupling between III-V and SiN sections (from completely isolated to totally coupled). Second is a consistent treatment of outcoupling, which enables the derivation of the laser equations to be more rigorous than possible with the customary one using quasi (non-normal) cavity modes. The description of the active medium follows multimode semiclassical laser theory, where electron-hole polarization dynamics accounts for both linear gain and carrier-induced refractive index change, as well as nonlinearities giving rise to saturation, mode competition and multiwave mixing. Quantum optical contributions are incorporated via a Langevin approach, with modifications to deal with the added complication of having a complex resonator geometry and coupling to free space.

Application of the approach identifies two physical processes underlying linewidth reduction. One is the gain clamping as is the case in single-cavity lasers. The second mechanism (unique to coupled cavities) is frequency locking of composite-cavity modes. Both mechanisms combine to describe details of emission spectra from below to above lasing threshold, such as deviation from a Lorentzian lineshape after the onset of lasing, as observed in experiments. Also described is coherence collapse because of incomplete frequency locking of composite-cavity modes. The periodic occurrences of coherence collapse may correspond to the experimentally observed periodic dips in intensity in heterogeneously-integrated III-V/SiN lasers.

Parametric studies suggest the possibility of laser linewidth reduction by six orders of magnitude from that of the high-$Q$ passive cavity. We presented results showing width (FWHM) of 0.6 Hz with SiN cavity $Q = 6 \times 10^7$. The parametric studies also indicate the role of the optical coupling between III-V and SiN sections. Preliminary results suggest that the interdependence of the optical coupling and passive-cavity $Q$ places a limit on the minimum intrinsic linewidth achievable through increasing the passive cavity $Q$-factor.

Lastly, the present formulation provides the basis of more detailed models for analyzing experiments and optimizing device engineering. An improvement is to use composite-cavity modes exactly for 2-D device geometries and model precisely the gratings and intermixing regions. In terms of laser theory, the strong signal treatment has room for greater rigor. Lastly, one should examine if there are consistency issues with our incorporation of quantum optical effects into a basically semiclassical laser theory. Given the goal of an analytical tool for engineering integrated III-V/SiN lasers, improvements should be introduced without increasing numerical complexity to the point of making parametric studies impractical. In fact, there is motivation for reducing the complexity of the present equations. The reason is so that bifurcation-continuation techniques may be used to systematically and extensively analyze the rich nonlinear dynamics found in closely identical, very weakly coupled oscillators (in our case rn and rw operation for $T_1 < 0.001$). [35] [36]. The next series of investigations will concentrate on high coherence lasers made with harmonic potential. Such complex semiconductor QW or QD lasers heterogeneously integrated with silicon photonics exhibit a large cavity quality factor, indicating their vast potential for optical radars, on-chip atomic clocks, and future coherent technologies.

**Acknowledgments**

This research was supported by Advanced Research Projects Agency-Energy (ARPA-E) No. DE-AR000067, the U.S. Department of Energy under Contract No. DE-AC04-94AL85000 and the American Institute for Manufacturing (AIM) Integrated Photonics. This work was performed, in part, at the Center for Integrated Nanotechnologies, an Office of Science User Facility operated for the U.S. Department of Energy (DOE) Office of Science. We thank C. Shen and L. Chang for discussions. The authors declare no conflicts of interest.